\documentclass[preprint]{revtex4}
\newcommand{\chem}[1]{\ensuremath{\mathsf{#1}}}       
\usepackage{graphicx}
\usepackage{longtable}

\begin{document}

\title{Calculation of chemical reaction energies using the AM05 density functional}
\author{Richard P. Muller and Ann E. Mattsson}
\affiliation{Multiscale Dynamic Material Modeling, Sandia National Laboratories, Albuquerque, NM}
\author{Curtis L. Janssen}
\affiliation{Scalable Computing Research and Development, Sandia National Laboratories, Livermore, CA}
\date{\today}

\begin{abstract}
We present results that compare the accuracy of the AM05 density functional \cite{Armiento:2005p5141, Mattsson:2008p5093} to a set of chemical reaction energies. The reactions were generated from the singlet species in the well-known G2 test suite \cite{Curtiss:1991p6667, Curtiss:1997p4252}. Our results show that, in general, the AM05 functional performs nearly as well as the other "pure" density functionals, but none of these perform as well as the hybrid B3LYP functional. These results are nonetheless encouraging because the AM05 functional arises from very simple assumptions, and does not require the calculation of the Hartree-Fock exchange integrals.
\end{abstract}

\preprint {2009-0460J} 
\preprint {JCC-09-0361.R1} 
\maketitle

\section{Introduction}
Density functional theory (DFT) \cite{HOHENBERG:1964p4075, KOHN:1965p4071} has become a central method in computational chemistry for understanding the energetics involved in structural changes to molecules and clusters. Conventional wisdom in the field states that hybrid density functionals, that is, functionals that combine the traditional DFT exchange with some amount of Hartree-Fock (HF) exchange \cite{BECKE:1993p3539}, are required to produce \emph{chemical accuracy} in molecules.  The use of hybrid functionals is often associated with  Becke \cite{BECKE:1993p3539} and the B3LYP functional, which combines HF exchange with the Becke 88 exchange functional \cite{Becke:1988p3531} and the Lee, Yang, and Parr \cite{LEE:1988p4435} correlation functional, and is the most widely-used DFT functional in chemistry. However, the use of hybrid exchange is not limited to B3LYP, and has been included in many other functionals \cite{Adamo:1999p2603, Xu:2005p3224, Zhao:2006p6627}.

The difficulty with the inclusion of any amount of HF exchange is that its computation is substantially more expensive. In non-periodic systems, molecules and clusters, this expense is normally small compared to the overall cost of the calculation. In solids, however, computation of the HF exchange represents a substantial portion of the overall computational expense. The situation presents something of a conundrum for those interested in chemical reactions in the condensed phase: the inclusion of the exchange is necessary for chemical accuracy, but renders the calculation intractable. Proper treatment of, say, heterogeneous catalysis, with density functional theory requires that the functional perform equally well for both molecules and solids. We therefore seek a pure density functional (one without HF exchange) that is capable of describing both molecules and solids.

The AM05 density functional \cite{Armiento:2005p5141, Mattsson:2008p5093} is one solution to this problem. AM05 is a simple density functional designed to describe electronic surfaces accurately. AM05 uses a subsystem functional approach \cite{KOHN:1998p8749, Armiento:2002p5139} to include the effect of electronic surfaces via the Airy gas \cite{KOHN:1998p8749}, while still retaining the physical consistency arising from the uniform electron gas approximation in the LDA. In recent investigations \cite{Mattsson:2008p5093} it has been shown to provide the same accuracy as hybrid functionals in calculations of bulk properties of solids, at a fraction of the computational cost. Our goal in the current paper is to understand how well this functional performs for chemical reactions.

As the spin-polarized version of AM05 is currently under development \cite{Mattsson:2009p10842}, we chose the subset of the G2 test suite \cite{Curtiss:1991p6667, Curtiss:1997p4252} having singlet ground states, computed the optimized structures for these compounds using the AM05, BLYP, PBE \cite{Perdew:1996p4232} and B3LYP density functionals. We then compute a wide range of chemical reaction energies using the total energies from these calculations. Our results suggest that, although none of the pure density functionals (i.e. BLYP, PBE, and AM05) perform as well as the hybrid functional we consider, AM05 performs roughly as well as the other two, suggesting that it can be used for accurate calculation of chemical reaction energetics is solids, where such numbers might otherwise be impossible to compute.

\section{Computational Method}
The AM05 functional is described at length in reference \cite{Armiento:2005p5141}. The Massively Parallel Quantum Chemistry (MPQC) program \cite{MPQC} was used for the calculation of the electronic structure of all species considered for each of the functionals. We used the 6-31G** double-$\zeta$-plus-polarization basis set in all calculations, and computed optimized geometries for each structure. The total energy of each structure was used to compute the reaction energy for the range of reactions being considered. In the computation of DFT energies, we do not include corrections due to zero-point vibrational energy. From these energies, we compute the reaction energy, the total energy difference between reactants and products, for a series of reactions of general interest. In the results section, below, we report the relative errors for these reaction energies with respect to the G2 energies for the B3LYP \cite{Becke:1988p3531,LEE:1988p4435,BECKE:1993p3539}, BLYP \cite{Becke:1988p3531,LEE:1988p4435}, PBE \cite{Perdew:1996p4232}, and AM05 \cite{Armiento:2005p5141, Mattsson:2008p5093} functionals. It is also worth noting that throughout this study we are comparing pure DFT total energy values to G2 values that have additional corrections, such as zero-point vibrational energy, that can be substantial for some of the reactions we consider.

\section{Results}
In this section we report the reaction energies for a variety of reactions constructed from the singlet molecules in the G2 test suite. Our aim in this comparison is to demonstrate that, although there can be quite substantial differences in the total energies that result from AM05 as compared to other density functionals, this functional can still be used to study the energy of a variety of chemical reactions. To facilitate the discussion in this section, we have grouped the reactions into broad categories related to the chemistry occurring in each case. Such a grouping is not intended to be canonical, as in some cases a reaction could be categorized in multiple groups, but rather as an aid to the discussion of the broad trends of the results. In each section we present the reactions and their energies with the different density functionals, with the results presented as mean absolute errors (MAE), with respect to the appropriate G2 energies.

\subsection{Hydrogenation reactions}
\begin{table}[htdp]
\caption{Errors in hydrogenation reactions relative to the G2 reference energies, in kcal/mol.}
\label{tab-hydrogenation}
\begin{tabular}{lrrrr}
Reaction			&				B3LYP &	BLYP &	PBE &	AM05\\ \hline
\chem{C_2H_4} + \chem{H_2} $\rightleftharpoons$ \chem{C_2H_6}	&			3.090&	0.807&	5.227&	7.077 \\
\chem{C_2H_2} + 2\ \chem{H_2} $\rightleftharpoons$ \chem{C_2H_6} &				8.557&	1.252&	12.744&	16.046\\
\chem{CO} + \chem{H_2} $\rightleftharpoons$ \chem{H_2CO}	&					4.717&	3.295&	8.944&	11.333\\
\chem{Li_2} + \chem{H_2} $\rightleftharpoons$ 2\ \chem{LiH}	&					2.169&	2.605&	1.633&	1.346\\
\chem{N_2} + 3\ \chem{H_2} $\rightleftharpoons$ 2\ \chem{NH_3}	&				1.217&	10.118&	2.194&	8.040\\
\chem{SiH_6} + \chem{H_2} $\rightleftharpoons$ 2\ \chem{SiH_4}	&				0.825&	0.151&	3.170&	3.640\\
\chem{N_2} + 2\ \chem{H_2} $\rightleftharpoons$ \chem{H_2NNH_2}&				3.136&	6.872&	8.112&	11.406\\
\chem{C_3H_6} (cyclopropene) +\chem{H_2} $\rightleftharpoons$ \chem{C_3H_8}	&	1.283&	3.828&	2.242&	3.956\\
\chem{CH_2}=\chem{C}=\chem{CH_2} +\chem{H_2} $\rightleftharpoons$ \chem{CH_3CH}=\chem{CH_2}	&	0.447&	5.228&	0.378&	1.547\\
\chem{CH_3CH}=\chem{CH_2}+\chem{H_2} $\rightleftharpoons$ propane		&			1.845&	2.226&	3.687&	5.225\\
2-butyne + 2\ \chem{H_2} $\rightleftharpoons$ \emph{trans}-butane		&			2.509&	5.174&	6.013&	8.882\\
butadiene + 2\ \chem{H_2} $\rightleftharpoons$ \emph{trans}-butane	&			2.270&	6.426&	5.706&	9.024\\
\chem{H_2} + \chem{CO_2} $\rightleftharpoons$ \chem{HCOOH}		&			1.055&	3.138&	1.152&	3.189\\
\chem{(CH_3)_2CO} + \chem{H_2} $\rightleftharpoons$ \chem{(CH_3)_2CHOH}	&	2.680&	7.267&	1.663&	0.268\\ \hline
MAE	 &							2.557&	4.170&	4.490&	6.498\\
\end{tabular}
\end{table}%

Table \ref{tab-hydrogenation} shows the errors in the various hydrogenation energies relative to the G2 reference energies. In general, the AM05 errors in the hydrogenation energies are similar to, but greater than, the errors produced by the other standard density functionals (BLYP and PBE), all of which are significantly greater than the errors produced by the hybrid functional B3LYP. For AM05, the most problematic reactions involve the hydrogenation of triple bonds such as in \chem{C_2H_2} or \chem{N_2}, a difficulty that is also reflected to a lesser degree in the other pure density functionals.

\subsection{Oxygenation reactions}
\begin{table}[htdp]
\caption{Errors in oxygenation reactions relative to the G2 reference energies, in kcal/mol.}
\label{tab-oxygenation}
\begin{tabular}{lrrrr}
Reaction			&				B3LYP &	BLYP &	PBE &	AM05\\ \hline
\chem{CH_4} + \chem{H_2O} $\rightleftharpoons$ \chem{CH_3OH} + \chem{H_2}	&	2.975	& 4.826 & 3.731 &	3.627\\
\chem{HOOH} + \chem{H_2} $\rightleftharpoons$ 2\ \chem{H_2O}	&	11.638 &	18.062 &	14.898 &	12.998\\
\chem{CO} + \chem{H_2O} $\rightleftharpoons$ \chem{CO_2} + \chem{H_2} &		10.540 &	15.447 &	17.742 &	19.681\\
\chem{C_2H_4} + \chem{HOOH} $\rightleftharpoons$ \chem{CHOCOH} + 2\ \chem{H_2} &		0.598 &	4.033 &	0.465 &	1.606\\
\chem{C_2H_6} + \chem{HOOH} $\rightleftharpoons$ \chem{CH_3CH_2OH} + \chem{H_2O} &		9.212 &	13.839 &	11.593 &	9.595\\
\chem{C_2H_4} + \chem{HOOH} $\rightleftharpoons$ \chem{C_2H_4O} + \chem{H_2O} &		5.350 &	9.626 &	3.379 &	0.302\\
\chem{CH_3SCH_3} + \chem{HOOH} $\rightleftharpoons$ \chem{(CH_3)_2SO} + \chem{H_2O} &		18.618 &	19.367 &	16.493 &	14.979\\
\chem{CH_3CHO} + \chem{H_2O} $\rightleftharpoons$ \chem{CH_3COOH} + \chem{H_2} &		6.189 &	7.805 &	8.646 &	10.131\\ \hline
MAE &	8.140 &	11.626 &	9.618 &	9.115\\
\end{tabular}
\end{table}%

Table \ref{tab-oxygenation} shows the errors in the various oxygenation energies relative to the G2 reference energies. Here the AM05 functional performs very well compared to the others, with a MAE of 9.115, close to the value of 8.140 achieved by B3LYP, and better than that of the other pure density functionals. The largest error for AM05 comes from the reaction $\chem{CO} + \chem{H_2O} \rightleftharpoons \chem{CO_2} + \chem{H_2}$, which potentially continues the trend with triple bonds giving difficulties for the AM05 functional relative to other chemical moieties.

\subsection{Nitrogen addition reactions}
\begin{table}[htdp]
\caption{Errors in nitrogen addition reactions relative to the G2 reference energies, in kcal/mol.}
\label{tab-nitrogen}
\begin{tabular}{lrrrr}
Reaction			&				B3LYP &	BLYP &	PBE &	AM05\\ \hline
\chem{CH_4} + \chem{NH_3} $\rightleftharpoons$ \chem{CH_3NH_2} + \chem{H_2}		& 0.892	& 2.070	& 1.982	& 2.307 \\
3 \chem{CH_4} + \chem{NH_3} $\rightleftharpoons$ \chem{(CH_3)_3N} + 3 \chem{H_2}		& 1.711	& 1.411	& 2.510	& 3.803 \\
\chem{NH_3} + 2 \chem{CH_4} $\rightleftharpoons$ \chem{(CH_3)_2NH} + 2 \chem{H_2}		& 0.538	& 2.740	& 3.069	& 3.866\\
\chem{CH_4} + \chem{HCN} $\rightleftharpoons$ \chem{CH_3CN} + \chem{H_2}		   	& 0.665	& 1.212	& 2.401	& 3.311\\
\chem{C_2H_6} + \chem{NH_3} $\rightleftharpoons$ \chem{CH_3CH_2NH_2} + \chem{H_2}	& 0.379	& 1.656	& 1.853	& 2.308\\
\chem{CH_2O} + \chem{NH_3} + \chem{CH_3CHO} $\rightleftharpoons$ \chem{C_5H_5N} + 3 \chem{H_2O}+\chem{H_2}		& 12.639	& 16.967	& 11.807	& 6.942\\
\chem{C_2H_2} + \chem{HCN} $\rightleftharpoons$ \chem{CH_2}=\chem{CHCN}		& 7.551	& 5.480	& 11.990	& 14.283\\
\chem{C_4H_4O} + \chem{NH_3} $\rightleftharpoons$ \chem{C_4H_5N} + \chem{H_2O}		& 4.609	& 6.653	& 4.314	& 3.092\\
\chem{CH_3CHO} + \chem{NH_3} $\rightleftharpoons$ \chem{CH_3CONH_2} + \chem{H_2}	& 16.451	& 15.435	& 14.546	& 13.549\\
2 \chem{HCN} $\rightleftharpoons$ \chem{NCCN} + \chem{H_2}				& 4.610	& 7.855	& 8.476	& 8.972\\
\chem{C_2H_4} + \chem{NH_3} $\rightleftharpoons$ \chem{C_2H_4NH} + \chem{H_2}		& 2.611	& 3.094	& 7.992	& 10.648\\ \hline
	MAE	& 4.787	& 5.870	& 6.449	& 6.644\\
\end{tabular}
\end{table}%

Table \ref{tab-nitrogen} shows the errors in the various nitrogen addition reactions, relative to the G2 reference energies. Here all functionals perform fairly well, with MAE values that span from 4.8--6.6 kcal/mol, with B3LYP being the best of these, and AM05 being the worst. It is difficult to draw too many conclusions about the trends in the AM05 behavior. The worst behavior is for the reaction \chem{C_2H_2} + \chem{HCN} $\rightleftharpoons$ \chem{CH_2}=\chem{CHCN}, and one would be tempted to attribute the poor behavior to the triple bond, which is problematic in other reactions studied in this paper, but the reaction \chem{CH_4} + \chem{HCN} $\rightleftharpoons$ \chem{CH_3CN} + \chem{H_2}, also containing the \chem{HCN} molecule, is one of the reactions on which AM05 performs best. AM05 also does poorly on the formation of acetamide, \chem{CH_3CHO} + \chem{NH_3} $\rightleftharpoons$ \chem{CH_3CONH_2} + \chem{H_2}, but as the other functionals all do similarly poorly, this is not a fault of AM05 alone.

\subsection{Halogenation reactions}
\begin{table}[htdp]
\caption{Errors in halogenation reactions relative to the G2 reference energies, in kcal/mol.}
\label{tab-halogenation}
\begin{tabular}{lrrrr}
Reaction			&				B3LYP &	BLYP &	PBE &	AM05\\ \hline
\chem{H_2} + \chem{Cl_2} $\rightleftharpoons$ 2\ \chem{HCl}	&	0.580 &	4.481 &	1.480 &	0.360\\
\chem{H_2} + \chem{F_2} $\rightleftharpoons$ 2\ \chem{FH} &		23.405 &	33.663 &	28.971 &	25.092\\
\chem{Li_2} + \chem{F_2} $\rightleftharpoons$ 2\ \chem{LiF} &		9.105 &	13.410 &	19.087 &	20.025\\
\chem{Na_2} + \chem{Cl_2} $\rightleftharpoons$ 2\ \chem{NaCl}	&	7.042 &	13.570 &	13.730 &	13.651\\
\chem{CH_4} + \chem{HCl} $\rightleftharpoons$ \chem{H_2} + \chem{CH_3Cl}	&	0.314 &	1.666 &	1.256 &	0.767\\
\chem{BF_3}+1.5\ \chem{Cl_2} $\rightleftharpoons$ \chem{BCl_3}+1.5\ \chem{F2}	&	4.596 &	10.478 &	17.640 &	15.730\\
\chem{AlF_3}+1.5\ \chem{Cl_2} $\rightleftharpoons$ \chem{AlCl_3}+1.5\ \chem{F_2}	 &	9.264 &	12.460 &	20.075 &	19.923\\
\chem{OF_2} + \chem{H_2} $\rightleftharpoons$ \chem{H_2O} + \chem{F_2}	&	10.376 &	18.948 &	16.657 &	15.702\\
\chem{SiH_4} + 2\ \chem{Cl_2} $\rightleftharpoons$ \chem{SiCl_4} + 2\ \chem{H_2}	&	19.143 &	25.110 &	16.551 &	14.393\\
\chem{SiH_4} + 2\ \chem{F_2} $\rightleftharpoons$ \chem{SiF_4} + 2\ \chem{H_2}	&	21.755 &	35.139 &	37.307 &	33.667\\
\chem{NH_3} + 1.5\ \chem{F_2} $\rightleftharpoons$ \chem{NF_3} + 1.5\ \chem{H_2}	&	1.436 &	6.433 &	7.725 &	11.038\\
\chem{PH_3} + 1.5\ \chem{F_2} $\rightleftharpoons$ \chem{PF_3} + 1.5\ \chem{H_2}	&	17.440 &	23.814 &	24.934 &	22.103\\
0.5\ \chem{Cl_2} + 1.5\ \chem{F_2} $\rightleftharpoons$ \chem{ClF_3}	&	13.499 &	8.868 &	3.407 &	0.587\\
\chem{C_2H_4} + 2\ \chem{F_2} $\rightleftharpoons$ \chem{C_2F_4} + 2\ \chem{H_2}	&	6.814 &	13.234 &	8.701 &	0.732\\
\chem{C_2H_4} + 2\ \chem{Cl_2} $\rightleftharpoons$ \chem{C_2Cl_4} + 2\ \chem{H_2}	&	8.313 &	8.717 &	0.772 &	5.688\\
\chem{CH_4} + \chem{F_2} $\rightleftharpoons$ \chem{CH_2F_2} + \chem{H_2}	&	8.458 &	12.965 &	12.123 &	8.913\\
\chem{CH_4} + 1.5\ \chem{F_2} $\rightleftharpoons$ \chem{CHF_3} + 1.5\ \chem{H_2}	&	9.812 &	16.821 &	14.100 &	8.001\\
\chem{CH_4} + \chem{Cl_2} $\rightleftharpoons$ \chem{CH_2Cl_2} + \chem{H_2}	&	1.777 &	2.353 &	0.347 &	1.555\\
\chem{CH_4} + 1.5\ \chem{Cl_2} $\rightleftharpoons$ \chem{CHCl_3} + \chem{H_2}	&	5.393 &	5.569 &	0.897 &	1.248\\
\chem{CH_3CN} + 1.5\ \chem{F_2} $\rightleftharpoons$ \chem{CF_3CN} + 1.5\ \chem{H_2}	&	7.316 &	12.613 &	10.114 &	4.164\\
\chem{C_2H_2} + \chem{HF} $\rightleftharpoons$ \chem{CH_2}=\chem{CHF}	 &	14.903 &	15.049	& 19.056&	20.759\\
\chem{C_2H_4} + \chem{HCl} $\rightleftharpoons$ \chem{C_2H_5Cl}	&	2.309&	0.169&	5.702&	7.201\\
\chem{C_2H_2} + \chem{HCl} $\rightleftharpoons$ \chem{CH_2}=\chem{CHCl}	&	5.646 &	3.971&	9.986&	11.702\\
\chem{H_2O} + \chem{Cl_2} $\rightleftharpoons$ \chem{HOCl} + \chem{HCl}	&	4.894&	6.023&	6.145&	6.190\\
\chem{CH_3CH}=\chem{CH_2} + \chem{HCl} $\rightleftharpoons$ \chem{CH_2CH_2CH_2Cl}	&	0.759&	1.920&	3.869&	5.057\\
2\ \chem{C_2H_4} + \chem{CCl_4} $\rightleftharpoons$ \chem{C_5H_8} + 2\ \chem{Cl_2}	&	9.398&	6.919&	13.189&	16.952\\
\chem{H_2CO} + \chem{F_2} $\rightleftharpoons$ \chem{CF_2O} + \chem{H_2}	&	2.352&	6.475&	2.348&	3.367\\
\chem{CH_3CHO} + \chem{HF} $\rightleftharpoons$ \chem{CH_3COF} + \chem{H_2}	&	9.197&	12.441&	11.810&	12.479\\ \hline
	MAE	& 8.403&	11.903 &	11.713&	10.966\\ 
\end{tabular}
\end{table}

Table \ref{tab-halogenation} shows the errors in the various halogenation energies, relative to the G2 reference energies. In general, this is another class of reactions in which the AM05 functional out performs the other pure density functionals, and performs nearly as well as the hybrid B3LYP functional. There are two important points to make for these reactions. First, although it is true that AM05 performs as well or better than the other density functionals for these reactions, in general the reaction energies for these reactions have larger errors than do the other reaction groups we consider here. Moreover, there are a number of systems, in particular those containing F atoms, which all of the functionals in general, and the AM05 in particular, have trouble with.

\subsection{Sulfur addition reactions}
\begin{table}[htdp]
\caption{Errors in sulfur addition reactions relative to the G2 reference energies, in kcal/mol.}
\label{tab-sulfur}
\begin{tabular}{lrrrr}
Reaction			&				B3LYP &	BLYP &	PBE &	AM05\\ \hline
\chem{CH_4} + \chem{SH_2} $\rightleftharpoons$ \chem{CH_3SH} + \chem{H_2} &		0.682&	0.302&	0.873&	0.730\\
\chem{CO_2} + \chem{CS_2} $\rightleftharpoons$ 2 \chem{COS}&		0.272&	0.752&	0.061&	0.367\\
\chem{C_2H_4} + \chem{SH_2} $\rightleftharpoons$ \chem{C_2H_4S} + \chem{H_2}&		0.449&	0.526&	5.266&	7.418\\
\chem{C_2H_6} + \chem{SH_2} $\rightleftharpoons$ \chem{CH_3CH_2SH} + \chem{H_2}	&	2.203&	1.238&	0.348&	0.395\\
\chem{C_4H_4O} + \chem{SH_2} $\rightleftharpoons$ \chem{C_4H_4S} + \chem{H_2O}&		11.591&	13.387&	10.263&	10.813\\ \hline
	MAE&	3.040&	3.241&	3.362&	3.945\\
\end{tabular}
\end{table}%

Table \ref{tab-sulfur} shows the errors in the various sulfur addition reactions, relative to the G2 reference energies. There are too few reactions here to draw very extensive conclusions about trends in the energetics. In general all of the functionals perform well for these reactions, with the AM05 functional yielding slightly higher errors relative to the G2 reference energies than the other reactions. Notable is that AM05 shows a higher error for the formation of \chem{C_2H_4S}, which has considerable ring strain, than do the other functionals. This might indicate a difficulty with ring strain, although the somewhat analogous reaction \chem{C_2H_4} + \chem{HOOH} $\rightleftharpoons$ \chem{C_2H_4O} + \chem{H_2O} forming oxirane rather than tioxirane is a reaction that AM05 has a much smaller error relative to the G2 reference energies than do the other functionals.

\subsection{Isomerization reactions}
\begin{table}[htdp]
\caption{Errors in isomerization reactions relative to the G2 reference energies, in kcal/mol.}
\label{tab-isom}
\begin{tabular}{lrrrr}
Reaction			&				B3LYP &	BLYP &	PBE &	AM05\\ \hline
allene $\rightleftharpoons$ propyne& 		4.109&	5.354&	5.265&	5.341\\
cyclopropene $\rightleftharpoons$ propyne	&	2.333&	1.842&	6.639&	8.740\\
\emph{trans}-butane $\rightleftharpoons$ isobutane	&	1.237&	1.332&	1.049&	1.017\\
propylene $\rightleftharpoons$ cyclopropane	&	0.046&	2.112&	3.238&	5.807\\
2-butyne $\rightleftharpoons$ butadiene	&	0.238&	1.253&	0.307&	0.142\\
2-butyne $\rightleftharpoons$ methylene cyclopropane	&	1.200&	0.469&	5.068&	7.684\\
2-butyne $\rightleftharpoons$ bicylobutane	&	0.984&	3.483&	6.567&	11.555\\
2-butyne $\rightleftharpoons$ cyclobutene	&	0.048&	1.442&	4.091&	6.940\\
nitrometh $\rightleftharpoons$ meth nitrite	&	0.086&	2.085&	1.311&	3.688\\
dimethamine $\rightleftharpoons$ transethylamine	&	1.156&	1.568&	0.878&	0.880\\
ethanol $\rightleftharpoons$ dimethyl ether	&	2.782&	4.017&	2.314&	1.804\\
ethanethiol $\rightleftharpoons$ dimethyl sulfide	&	0.648&	1.111&	0.935&	0.526\\ \hline
	MAE&	1.239&	2.172&	3.138&	4.510\\
\end{tabular}
\end{table}%

Table \ref{tab-isom} shows the errors in the various isomerization reactions, relative to the G2 reference energies.  In general, the performance of all the functionals is quite good. Here the AM05 functional does worse than either of the other pure functionals. We note, in particular, that the trend of the errors being highest for reactions that either have ring strain or triple bonds continues with these reactions as well.

\subsection{Carbon or silicon addition reactions}
\begin{table}[htdp]
\caption{Errors in carbon or silicon addition reactions relative to the G2 reference energies, in kcal/mol.}
\label{tab-carbon}
\begin{tabular}{lrrrr}
Reaction			&				B3LYP &	BLYP &	PBE &	AM05\\ \hline
\chem{CH_4} + \chem{SiH_4} $\rightleftharpoons$ \chem{CH_3SiH_3} + \chem{H_2}	&	2.572&	2.159&	0.816&	0.636 \\
\chem{HCOOH} + \chem{CH_4} $\rightleftharpoons$ \chem{HCOOCH_3} + \chem{H_2}	&	0.311&	1.248&	0.259&	0.033\\
\chem{CH_4} + \chem{CO} $\rightleftharpoons$ \chem{CH_2CO} + \chem{H_2}	&	4.779&	8.435&	11.136&	13.789\\
\chem{CH_4} + \chem{CO} $\rightleftharpoons$ \chem{CH_3CHO}	&	3.829&	2.969&	9.871&	13.148\\
2 \chem{CH_3OH} $\rightleftharpoons$ \chem{CH_3OCH_3} + \chem{H_2O} &		1.739&	1.897&	1.504&	1.369\\
\chem{CH_3OH} + \chem{CH_3CH_2OH} $\rightleftharpoons$ \chem{C_2H_5OCH_3}+\chem{H_2O}	&	2.039&	2.182&	1.791&	1.655\\
3 \chem{C_2H_2} $\rightleftharpoons$ \chem{C_6H_6}	&	18.150&	9.755&	30.222&	38.383\\  \hline
	MAE	&4.774&	4.092&	7.943&	9.859\\
\end{tabular}
\end{table}%

Table \ref{tab-carbon} shows the errors in the various carbon and silicon addition reactions, relative to the G2 reference energies. Several species in this category reveal problems for the AM05 functional. The reactions containing \chem{CO} show the problem that AM05 has shown in other reactions with triple bonds. But of greatest concern is the results for benzene formation from acetylene, which shows a error in the reaction energy of 38.4 kcal/mol. The latter reaction also poses significant trouble for the PBE and B3LYP functionals. It is also worth restating that throughout this study we are comparing pure DFT total energy values to G2 values that have additional corrections, such as zero-point vibrational energy, that are substantial for some of these reactions.

\section{Conclusion}
\begin{table}[htdp]
\caption{Summary of the results presented in tables \ref{tab-hydrogenation}--\ref{tab-carbon}. MAE, in kcal/mol, with respect to the G2 energies.}
\label{tab-summary}
\begin{tabular}{lrrrr}
Reaction Class			&	B3LYP &	BLYP   &	PBE &	AM05\\ \hline
Hydrogenation Reactions &	2.557&	4.170  &	4.490&	6.498\\
Oxygenation Reactions &		8.140 &	11.626 &	9.618 &	9.115\\
Nitrogen Addition Reactions & 4.787 &      5.870  & 6.449 & 6.644 \\
Halogenation Reactions	& 	8.403&	11.903 &	11.713 &	10.966\\ 
Sulfur Addition Reactions &	3.040&	3.241   &	3.362&	3.945\\
Isomerization Reactions  &	1.239&	2.172   &	3.138&	4.510\\
Carbon/Silicon Addition Reactions &4.774&4.092&	7.943&	9.859\\ \hline
All Reactions & 5.322 & 7.296 & 7.633 & 8.081
\end{tabular}
\end{table}
None of the pure density functionals achieves the accuracy, compared to the G2 data, that the hybrid B3LYP functional achieves. The data presented in tables \ref{tab-hydrogenation}--\ref{tab-carbon} are summarized in table \ref{tab-summary}. We find, for the set of chemistry we consider here, that the B3LYP functional has only 5.32 kcal/mol mean absolute error with respect to the G2 data. In contrast, the PBE functional has 7.63 kcal/mol, the BLYP functional 7.29 kcal/mol, and the AM05 functional has 8.08 kcal/mol mean absolute error as compared to the G2 data. 

It is dangerous to draw too many conclusions from the grouped reaction energies as displayed in table \ref{tab-summary} about why different functionals perform well in some cases and poorly in others. In general, the AM05 functional performs most poorly in systems with ring strain and triple bonds. Because the AM05 functional is based on two model systems that are infinite, the uniform electron gas and the Airy gas, there are problems in systems with confined densities, which will be addressed in the successors to the AM05 functional.

In spite of the fact that the AM05 functional performs slightly worse with respect to the G2 data than the PBE and BLYP functionals, we are heartened by the results here. In the derivation of the AM05 functional, no adjustable parameters, and no knowledge of chemistry or even molecules were used. The lack of adjustable parameters, and the lack of a need for the Hartree-Fock exchange to be computed are shared by the BLYP and PBE functionals. The superior performance of the AM05 functional in solids, which was one of the motivating factors in this study, coupled with the good performance in chemical reaction energies, suggests that AM05 is a good choice when considering reaction energies in the solid phase. Furthermore, the AM05 functional has been constructed in a very different way from other functionals, with no information about chemistry or molecules used in its design. The fact that this functional performs nearly as well as the standard functionals is encouraging, because it suggests new directions and new ideas that may be used for the development of further improved functionals.

\section{Acknowledgments} 
Sandia is a multiprogram laboratory operated by the Sandia Corporation, a Lockheed Martin Company, for the United States Department of Energy under Contract No. DE-AC04-94AL85000.

\bibliographystyle{unsrt}
\bibliography{am05-chemistry}

\end{document}